\documentclass[7pt]{article}
\usepackage{amsmath,amssymb,graphicx}
\title{Thin-shell wormholes supported by ordinary matter in  Einstein--Gauss--Bonnet gravity}
\author{Mart\'{\i}n G. Richarte and Claudio Simeone\thanks{E-mail: csimeone@df.uba.ar. }\\
{\small  Departamento de F\'{\i}sica, Facultad de Ciencias Exactas y 
Naturales, Universidad de Buenos Aires}
   \\ 
{\small Ciudad Universitaria, Pabell\' on I, 1428, 
Buenos Aires, Argentina}} 

\begin{document}
\maketitle

\begin{abstract}

\noindent The generalized Darmois--Israel  formalism for Einstein--Gauss--Bonnet theory is applied  to  construct thin-shell Lorentzian wormholes with spherical symmetry.  We calculate the energy localized on the shell, and we find that for certain values of the parameters wormholes could be supported by matter not violating the energy conditions.

\end{abstract}

\section{Introduction}

For spacetime dimension $D\geq 5$  the  Einstein--Hilbert action of  gravity  admits quadratic corrections constructed  from coordinate-invariant tensors which scale as fourth derivatives of the metric. In particular, when $D=5$  the most general theory leading to  second order equations for the metric is the so-called Einstein--Gauss--Bonnet theory or Lovelock theory up to second order. This  class of model for higher dimensional gravity has been widely studied, in particular because it can be obtained in the low energy limit of  string  theory \cite{1}. For spacetime dimensions $D<5$ the Gauss--Bonnet terms in the action represent a topological invariant. 

 The equations of gravitation admit  solutions, known as Lorentzian wormholes, which connect two regions of the same universe (or of two universes) by a throat, which is a minimal area surface \cite{motho,book}. Such kind of geometries would present some features of particular interest, as for example the possibility of time travel (see Refs.  \cite{morris-novikov}). But a central objection against the actual existence of wormholes is that  in Einstein gravity the  flare-out condition \cite{hovis1} to be  satisfied at the throat  requires the presence of exotic matter, that is, matter violating the energy conditions \cite{book}. In this sense,  thin-shell wormholes have the advantage that the exotic matter would be located only at the shell.

 However, it has recently been shown  \cite{gra-wi} that in pure Gauss--Bonnet gravity exotic matter is no needed  for wormholes to exist; in fact, they could exist even with no matter (see also Refs. \cite{7,maeda}).  In this work we  thus study thin-shell wormholes in Einstein--Gauss--Bonnet gravity.  We focus in the amount of matter necessary for supporting the wormholes, without analyzing the microphysics explaining this matter. Differing from the approach in the related work Ref. \cite{grg06}, where the 
Gauss--Bonnet terms were treated as an effective source for the Einstein's field equations,  here 
  the Gauss--Bonnet contribution is treated  as an essentially  geometrical object. This requires a generalization \cite{4} of the Darmois--Israel formalism \cite{3} for thin shells, but provides a better physical understanding. In particular, we  show that for certain values of the parameters,  thin-shell wormholes could be supported by matter not violating the energy conditions.

\section{Spherically symmetric geometry}

We start from  the action for  Einstein--Maxwell--Gauss--Bonnet theory in a five-dimensional manifold ${\cal M}_5$ with cosmological constant  $\Lambda$ and   Maxwell field \cite{2}:  
\begin{equation}
S =\kappa\int_{{\cal M}_5} d^{5}x \sqrt{|g|}\left[ R-2\Lambda + \alpha R^{2}_{GB}-\frac{1}{4}F^{\mu\nu}F_{\mu\nu}\right] ,\label{action}
\end{equation}
where $\kappa$ is related with the Newton constant, $ R^{2}_{GB}=R^{2}-4R^{ab}R_{ab}+R^{abcd}R_{abcd}$ is the Gauss--Bonnet term, and  $\alpha$ is the  Gauss--Bonnet coupling constant. The Gauss--Bonnet constant introduces a scale $l^2_{GB}\propto |\alpha|$ in the theory which physically represents a short-distance correction to general relativity. Within string theory, in five dimensions $\alpha$ would be of order the string mass scale; but in a more general framework  $\alpha$ can be considered as an arbitrary real number with the appropriate dimensions. 

The field equations resulting from the action (\ref{action}) are
\begin{equation}
G_{ab}+2\alpha H_{ab}+ \Lambda g_{ab}=\kappa^{2}T_{ab}\label{field}
\end{equation}
where $H_{ab}$ is the second order Lovelock tensor  and $T_{ab}$ is the usual electromagnetic energy-momentun tensor:
\begin{equation} 
H_{ab}=RR_{ab}-2R_{ac}R^{c}_{b}-2R^{cd}R_{acbd}+R_{a}^{cde}R_{bcde}-\frac{1}{4}g_{ab}(R^{2}-4R^{cd}R_{cd}+R^{cdeq}R_{cdeq}),
\end{equation}
\begin{equation}
T_{ab}=F_{ac}F^{c}_{b}-\frac{1}{4}g_{ab}F_{cd}F^{cd}.
\end{equation}
Equations (\ref{field}) admit a spherically symmetric  solution given by \cite{2}:
\begin{equation}
{ds}^{2}=-f(r)\,dt^{2}+f^{-1}(r)\,dr^{2}+r^{2}d\Omega^{2}_{3},\label{metric}
\end{equation}
\begin{equation}
f(r)=1+\frac{r^{2}}{4\alpha}\left[1-\sqrt{1+\frac{16M\alpha}{\pi r^{4}}-\frac{8Q^{2}\alpha}{3 r^{6}}+\frac{4 \Lambda \alpha}{3}}\right].\label{f}
\end{equation}
 It is easy to check that in the limit $\alpha \rightarrow 0$ the five-dimensional Einstein--Maxwell solution with cosmological constant is recovered. Further, in this limit and  for $\Lambda=0$ the five-dimensional Reissner--N\"{o}rdstrom metric is obtained, so  $M>0$ and $Q$ can be indentified with the mass and the charge of the system. For $\alpha\neq 0$, there is, in principle, a minimun value of the radial coordinate $r_{min}$ such that the function under the square root in (\ref{f}) is positive so the metric (\ref{metric}) is well defined. The geometry has a curvature singularity at the surface defined by $r=r_{min}$. Depending on the values of the parameters $(M,\alpha,Q,\Lambda)$, this singular  surface can be surrounded by an event horizon with a radius $r_{hor}$, so the metric (\ref{metric}) represents a black hole; if no event horizont exists, it presents a naked singularity.
Here we will focus in the case of null cosmological constant; the singularity in the metric from which we start would present a singularity at $r_{min}$ given by the greatest real and positive solution of the equation
\begin{equation}
r^{6}+\frac{16M\alpha}{\pi} r^{2}-\frac{8Q^{2}\alpha}{3}=0.\label{r6}
\end{equation}
If Eq. (\ref{r6}) has no real positive solutions we have $r_{min}=0$, where the metric diverges. On the other hand, when horizons exist their radii are given by 
\begin{equation}
r_{\pm}=\sqrt{\frac{M}{\pi}-\alpha \pm\left[\left(\frac{M}{\pi}-\alpha\right)^{2}-\frac{Q^{2}}{3}\right]^{1/2}}.
\end{equation}
 So there exits a critical value of the charge:
\begin{equation}
\left|Q_{c}\right| =\sqrt{3}\left|\frac{M}{\pi}-\alpha\right|
\end{equation}
such that if $\left|Q\right|<\left|Q_{c}\right|$ there would exist two horizons, if $\left|Q\right|= \left|Q_{c}\right|$ there would exist  only one (degenerate) horizon, and if $\left|Q\right|>\left|Q_{c}\right|$ there are no horizons. 
The event horizont would be placed at $r_{hor}=r_{+}$, and $r_{-}$ would correspond to the inner horizon. If  $r_{min} <r_{hor}$  the singularity would be shielded by the event horizon, but if  $r_{min}\geq r_{hor}$ we would have a naked singularity. 

\section{Thin-shell wormhole construction}

Starting from the metric given by (\ref{metric}) and (\ref{f})  with $\Lambda =0$ we build a spherically symmetric thin-shell wormhole in the Einstein--Maxwell--Gauss--Bonnet theory. We take two copies of the  spacetime and remove from each manifold the five-dimensional regions described by 
\begin{equation}
{\cal M}_{1,2}=\left\{X/r_{1,2}\leq b\right\},
\end{equation} 
 where $b$ is chosen to include possible singularities or horizons within the regions  ${\cal M}_{1,2}$. The resulting manifolds have boundaries given by the timelike hypersurfaces 
\begin{equation}
\Sigma_{1,2}=\left\{X/r_{1,2} = b\right\}.
\end{equation}
 Then we identify these two timelike hypersurfaces to obtain a geodesically complete new manifold $\cal M$ with a matter shell at the surface $r=b$, where the throat of the wormhole is located. This manifold without singularities or horizons is constituted by two asymptotically flat regions.
To study this type of wormhole we apply the  Darmois--Israel formalism generalized \cite{4} to the case of Einstein-- Gauss--Bonnet theory. We can introduce the coordinates $\xi^{i}=(\tau,\theta,\chi,\varphi)$ in $\Sigma$, with $\tau$ the proper time on the throat. Though  we will focus mainly in static configurations, in the general case we could  allow the radius of the throat be a function of the proper time, so that the  boundary hypersurface reads:
\begin{equation}
\Sigma: {\cal F}(r,\tau)=r-b(\tau)=0.
\end{equation}
The  field equations projected on the shell $\Sigma$ (generalized Darmois--Israel conditions)  are \cite{4} 
\begin{equation}
2\langle K_{ab}-K h_{ab}\rangle + 4\alpha \langle 3J_{ab}-Jh_{ab}+2P_{acdb}K^{cd}\rangle =-\kappa^{2}S_{ab},
\end{equation}
where the $\langle \cdot\rangle$ stands for  the jump of a given quantity across the  hypersurface $\Sigma$.
The extrinsic curvature $K_{ab}$, the divergence-free part of the Riemann tensor $P_{abcd}$ and the tensor $J_{ab}$  are defined as follows:
\begin{equation}
{K}^{\pm}_{ab}=-n^{\pm}_{c}\left(\frac{\partial^{2}X^{c}}{\partial\xi^{a}\partial\xi^{b}}+\Gamma^{c}_{de}\frac{\partial X^{d}}{\partial\xi^{a}}\frac{\partial X^{e}}{\partial\xi^{b}}\right)_{r=b},
\end{equation} 
\begin{equation}
P_{abcd}=R_{abcd}+(R_{bc}h_{da}- R_{bd}h_{ca})-(R_{ac}h_{db}- R_{ad}h_{cb})+\frac{1}{2}R(h_{ac}h_{db}-h_{ad}h_{cb}),
\end{equation}
\begin{equation}
J_{ab}=\frac{1}{3}\left[2KK_{ac}K^{c}_{b}+K_{cd}K^{cd}K_{ab}-2K_{ac}K^{cd}K_{db}-K^{2}K_{ab}\right].
\end{equation}
The tensor $P_{abcd}$  is calculated with the induced metric $h_{ab}=g_{ab}-n_{a}n_{b}$ (see \cite{4}). After some algebraic manipulation, the non-null components  $S_{a}^{b}$ of the energy-momentun tensor of the shell are obtained as
\begin{equation}
S^{\tau}_{\tau}=\frac{1}{8\pi}\left[ 6\frac{\Delta}{b}-2\alpha\left(4\frac{\Delta^{3}}{b^{3}}-12(1+{\dot{b}}^{2})\frac{\Delta}{b^{3}}\right)\right],\label{sigma}
\end{equation}
\begin{equation}
S^{\theta}_{\theta}=S^{\varphi}_{\varphi}=S^{\chi}_{\chi}=\frac{1}{8\pi}\left[ 4\frac{\Delta}{b}+2\ell(b)\frac{1}{\Delta}-2\alpha\left(4\ell(b)\frac{\Delta}{b^{2}} -4(1+{\dot{b}}^{2})\frac{\ell(b)}{b^{2}\Delta }-8\frac{\ddot{b}\Delta}{b^2}\right)\right],\label{p}
\end{equation}
where $\ell(b)=\ddot{b}+f^{'}(b)/2$ and $\Delta=\sqrt{{\dot{b}}^{2}+f(b)}$;  
the dot means a derivative with respect to the proper time and the prime with  respect to $b$. 
From these equations  we read the  energy density $\sigma=-S^{\tau}_{\tau}$ and  the tranverse pressure $p=S^{\theta}_{\theta}=S^{\chi}_{\chi} =S^{\varphi}_{\varphi}$ in terms of the throat radius $b(\tau)$, first and second derivatives of $b(\tau)$ and the function $f(b)$ wich depends on the parameters of the system.
If we take $\alpha\rightarrow 0$ in both equations  (\ref{sigma}) and (\ref{p}) we recover the expression for the energy density $\sigma$ and  pressure $p$ found in Ref. \cite{grg06} with the standard Lanczos equation for the shell. Furthermore,  Taylor expanding up to zeroth order in $\alpha$  we recover the expressions for the  five-dimensional Schwarzschild and Reissner--N\"{o}rdstrom cases.
Starting from Eqs. (\ref{sigma}) and (\ref{p}), in the next Section we will show  that for certain values of the parameters, ordinary matter could support thin-shell wormholes in  Einstein--Gauss--Bonnet theory.

\section{Matter supporting the wormholes; discussion}

 Motivated by the results within pure Gauss--Bonnet gravity (i.e. without Einstein term) in Ref. \cite{gra-wi}, here we evaluate the amount of exotic matter and the energy conditions, following the approach presented above in which the Gauss--Bonnet term is treated as a geometrical contribution in the field equations. Coming this contribution from the curvature tensor, this approach is clearly the most suitable to give a precise meaning to the characterization of matter supporting the wormhole. As we shall see, the results will considerably differ from those in Ref. \cite{grg06}, where the Gauss--Bonnet term was treated as an effective source for the Einstein's field equations. 

The {\it weak energy condition} (WEC) states that for any timelike  vector $U^{\mu}$ it must be $T_{\mu\nu}U^{\mu}U^{\nu}\geq0$; the WEC also implies, by continuity, the {\it null energy condition} (NEC), which means that for any null vector $k^{\mu}$ it mus be $T_{\mu\nu}k^{\mu}k^{\nu}\geq0$ \cite{book}. In an orthonormal basis the WEC reads $\rho\geq0$, $\rho+p_{l}\geq0\  \forall\, l$, while the NEC takes the form $\rho+p_{l}\geq0 \ \forall\, l$. In the case of  thin-shell wormholes  the radial pressure $p_{r}$ is zero, while within Einstein gravity or even with the inclusion of a Gauss--Bonnet term in the way proposed in \cite{grg06}, the surface energy density must fulfill  $\sigma<0$, so that both energy conditions would be  violated. The sign of $\sigma+p_{t}$  where $p_{t}$ is the transverse pressure  is not fixed, but it depends on the values of the parameters of the system.

In what follows we restrict to static configurations. The surface  energy density $\sigma_{0}$ and the transverse pressure $p_{0}$ for a static  configuration ($b=b_0$, $\dot{b}=0$, $\ddot{b}=0$) are given by
\begin{equation}
\sigma_{0}=-\frac{1}{8\pi}\left[ 6\frac{\sqrt{f(b_{0})}}{b_{0}}-2\alpha\sqrt{f(b_{0})}\left(4\frac{f(b_{0})}{b^{3}_{0}}-\frac{12}{b^{3}_{0}}\right)\right],\label{s0}
\end{equation}
\begin{equation}
p_{0}=\frac{1}{8\pi}\left[ 4\frac{\sqrt{f(b_{0})}}{b_{0}}+\frac{f^{'}(b_{0})}{\sqrt{f(b_{0})}}-2\alpha\left(2f^{'}(b_{0})\frac{\sqrt{f(b_{0})}}{b^{2}_{0}} -2\frac{f^{'}(b_{0})}{b^{2}_{0}\sqrt{f(b_{0})} }\right)\right].\label{p0}
\end{equation}
Note that the sign of the surface energy density is, in principle, not fixed. The most usual choice  for quantifying the  amount of exotic matter in a Lorentzian wormhole is the integral \cite{nandi}:
\begin{equation}
\Omega= \int (\rho + p_{r})\sqrt{|g|}\,d^{4}x.
\end{equation}
We can introduce a new radial coordinate ${\cal R}=\pm(r-b_{0})$ with $\pm$ corresponding to each side of the shell. Then,  
because  in our construction the energy density is located on the surface, we can write $\rho=\delta({\cal R})\,\sigma_{0}$, and because the shell does not exert radial pressure  the amount of exotic matter reads
\begin{equation}
\Omega=\int^{2\pi}_{0} \int^{\pi}_{0}\int^{\pi}_{0}\int^{+\infty}_{-\infty}\delta({\cal R})\,\sigma_{0} \sqrt{|g|}\, d{\cal R}\,d\theta\,d\chi\ d\varphi
      =2\pi^{2}  b_{0}^3 \sigma_{0}.
\end{equation}
Replacing the explicit form of $\sigma_{0}$ and $g$, we obtain the exotic matter amount as a function of the parameters  that characterize the configurations:
\begin{equation}
{\Omega}= -\frac{3}{2}\pi b^{2}_{0}\sqrt{f(b_{0})} +2\pi\alpha\sqrt{f(b_{0})}\left[f(b_{0})-3)\right],
\end{equation}
where $f$ is given by (\ref{f}). For  $\Lambda=0$, in the limit $\alpha\rightarrow 0$ and Taylor expanding up to zeroth order we obtain the exotic matter for  the Reissner--N\"{o}rdstrom ($Q\neq 0$) and Schwarzschild ($Q= 0$) geometries.

For $\alpha\neq 0$  we now find that there exist positive contributions to $\Omega$; these come from the different signs in the expression (\ref{s0}) for the surface energy density, because $\Omega$ is proportional to $\sigma_0$. We stress that this would not be possible if the standard Darmois--Israel formalism was applied, treating the Gauss--Bonnet contribution as an effective energy-momentum tensor, because this leads to $\sigma_0\sim - \sqrt{f(b_0)}/b_0$ \cite{grg06}. Now, once the explicit form of the function $f$ (with $\Lambda=0$) is introduced, the condition $\sigma_0>0$ leads to 
\begin{equation}
-8\alpha-2b_0^2-b_0^2\sqrt{1+\frac{16M\alpha}{\pi b_0^4}-\frac{8Q^2\alpha}{3b_0^6}}>0,\label{s+}
\end{equation} 
which can hold only for $\alpha <0$.
The subsequent  analysis is simplified  by considering the  case $Q=Q_c, \,\Lambda=0$; then there would be at most only one horizon in the original manifold, its radius being independent of the charge. But for this  charge  it can be shown that, for values of $\alpha$ such that the horizon exists, it is not possible to fulfil Eq. (\ref{s+}) for any wormhole radius larger than $r_{hor}$; the reason is that the horizon exists only for $\alpha > -M/(3\pi)$, which is not compatible with condition (\ref{s+}) if $b_0$ is to be larger than the corresponding horizon radius. Instead, a simple numerical analysis shows that for $\alpha$ slightly below  $-M/(3\pi)$ both the singularity at $r\neq 0$ and the horizon dissapear in the original manifold, so that the only condition to be fulfilled  is that given by Eq. (\ref{s+}). And for $\alpha <0$ it is always possible to choose $b_0$ such that this indeed happens, so  that   the existence of  thin-shell wormholes  is compatible with a positive surface energy density \footnote{It is not difficult to see that for $Q$ slightly below $Q_c$ the same happens for larger  $|\alpha|$.}.

In figures 1 and  2 we show the  amount $\Omega$ as a function of the wormhole radius  for this  relatively large  value of $|\alpha|$ (that is, $\pi |\alpha|$ of order $M$); though this would imply microscopic configurations or a scenario different from that suggested by present day observation, the  analysis shows that this is the most interesting situation. Besides the fact that $\Omega$ results to be smaller when calculated by treating the Gauss--Bonnet contribution as a geometric object than in the case that it was treated as an effective energy-momentum tensor, this amount  is  smaller than which would be necessary in the five-dimensional  pure Reissner--N\"{o}rdstrom case (see Fig. 1). However, the central, remarkable, result is that  we have a region with $\Omega >0$ (see Fig. 2), corresponding to  $\sigma_0>0$; and that besides, from  Eqs. (\ref{s0}) and (\ref{p0}) we have  
$\sigma_0+p_0=-(b_0/3)\,d\sigma_0/db_0$,
which shows that for  wormhole radii such that $\sigma_0 >0$ and ${d\sigma_0}/{db_0}<0$ ($r_{wh}$ within the maximun and the zero of $\sigma_0$ in Fig. 3) both the WEC and the NEC are satisfied. Thus, in the picture providing a clear meaning to matter in the shell, in Einstein--Gauss--Bonnet gravity the violation of the  energy conditions could be avoided, and  wormholes could be supported by ordinary matter. 

\begin{figure}[htp]
\centering
\includegraphics[height=8cm, width=12cm]{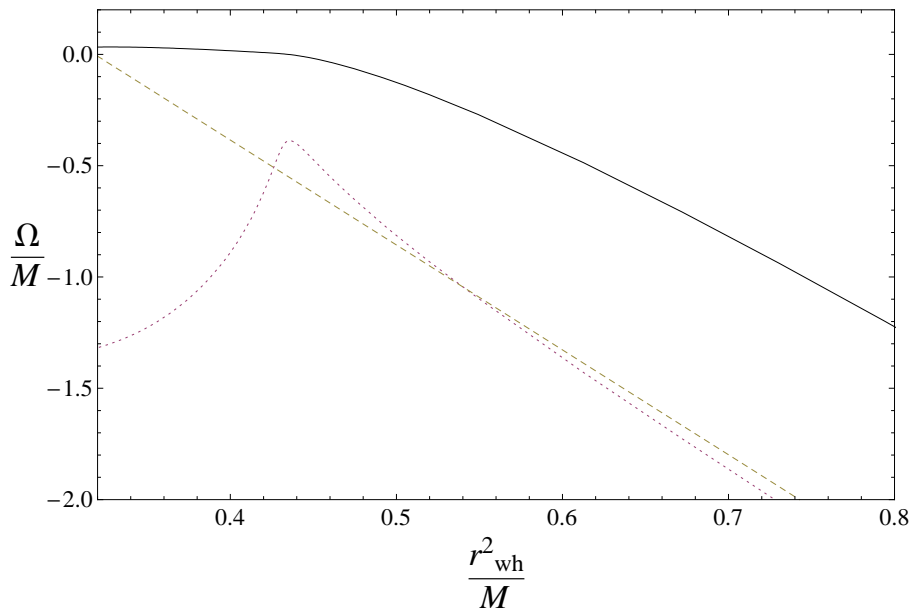}
\caption{The  amount $\Omega$   is shown as a function of $r^2_{wh}/M$, for $Q=Q_c$ and  $\alpha=-0.11 M$. The  dashed line  corresponds to the  five-dimensional Reissner--N\"{o}rdstrom case, the dotted line corresponds to considering  the Gauss--Bonnet term as a kind of effective source for the field equations, and the  solid  line shows the result obtained here with the generalized Darmois--Israel formalism for Einstein--Gauss--Bonnet theory.}
\end{figure}
\begin{figure}[htp]
\centering
\includegraphics[height=8.cm, width=8.5cm]{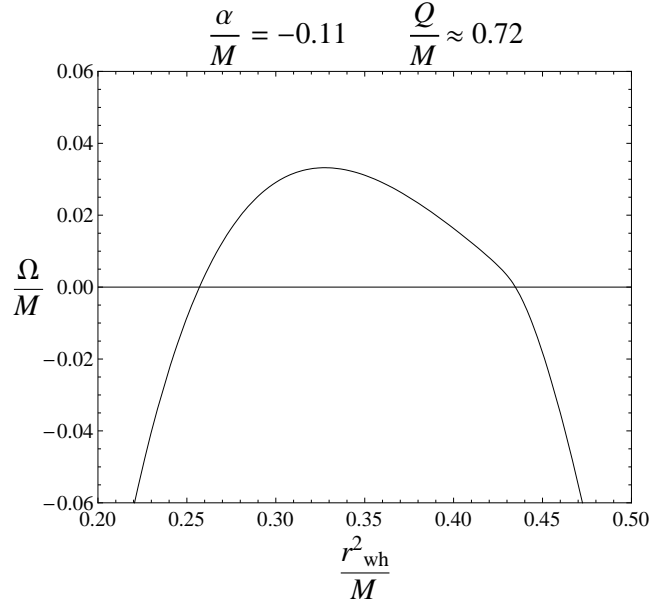}
\caption{The  amount $\Omega$  is shown as a function of $r^2_{wh}/M$, for $Q=Q_c$ and  $\alpha=-0.11 M$.  The plot shows the result obtained here with the generalized Darmois--Israel formalism for Einstein--Gauss--Bonnet theory.}
\end{figure}
\begin{figure}[htp]
\centering
\includegraphics[height=8cm, width=8cm]{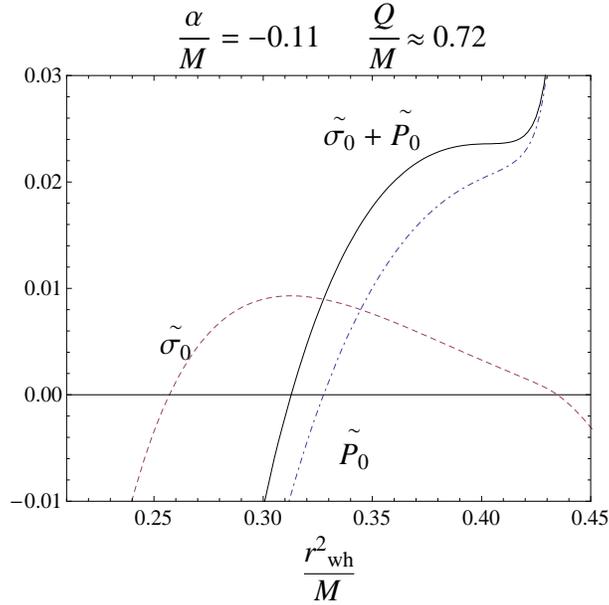}
\caption{Energy conditions: the dashed line shows $\tilde \sigma_0=\sqrt{M}\sigma_0$, the dashed-dotted line shows $\tilde p_0=\sqrt{M}p_0$ and the solid line shows the sum $\tilde \sigma_0+\tilde p_0$.}
\end{figure}

\section*{Acknowledgments}
We wish to thank  M. Thibeault and E. Eiroa for important suggestions. 
This work was supported by  Universidad de Buenos Aires and CONICET.

\end{document}